\documentclass[12pt]{article}
\usepackage{graphicx}
\usepackage{lineno, blindtext}
\usepackage{wrapfig}
\usepackage{siunitx}
\usepackage{xcolor}
\usepackage{subcaption}
\usepackage{blindtext}
\usepackage{hyperref}   
\usepackage{lipsum}

\usepackage[a4paper]{geometry} 

\def\Title#1{\begin{center} {\Large #1 } \end{center}}
\def\Author#1{\begin{center}{ \sc #1} \end{center}}
\def\Address#1{\begin{center}{ \it #1} \end{center}}

\newenvironment{Abstract}{\begin{quotation}  }{\end{quotation}}
\newenvironment{Presented}{\begin{quotation} \begin{center} 
             PRESENTED AT\end{center}
      \begin{center}\begin{large}}{\end{large}\end{center} \end{quotation}}

\begin{document}
\begin{titlepage}
\Title{Transverse Single-Spin Asymmetry for Inclusive and Diffractive Electromagnetic Jets at Forward Rapidity in $p^{\uparrow}$+p Collisions at $\sqrt{s} = 200$ GeV and $510$ GeV at STAR}
\vfill
\Author{Xilin Liang, for the STAR Collaboration}
\Address{University of California, Riverside}
\vfill
\begin{Abstract}
There have been numerous attempts, in the last decades, to understand the origin of the unexpectedly large transverse single-spin asymmetry ($A_{N}$) observed in inclusive hadron productions at forward rapidities in transversely polarized $p^{\uparrow}$+$p$ collisions at different center-of-mass energies ($\sqrt{s}$). The current theoretical frameworks aimed at explaining this puzzle include the twist-3 contributions in the collinear factorization framework, as well as the transverse-momentum-dependent contributions from the initial-state quark and gluon Sivers functions, and/or final-state Collins fragmentation functions. Besides, there are indications that the diffractive processes may contribute to the large $A_{N}$. We present the detailed investigations into the $A_{N}$ for electromagnetic jets (EM-jets) produced in inclusive processes using the Forward Meson Spectrometer with transversely polarized $p^{\uparrow}$+ $p$ data at $\sqrt{s} =$ 200 GeV collected in 2015 at STAR.  We observe a negative value for the $A_{N}$ of EM-jets in diffractive processes. This finding shows a different sign for $A_{N}$ in inclusive processes and needs further theoretical input in order to be understood. Finally, we present the statistical projections of the $A_{N}$ for inclusive and diffractive EM-jets utilizing $p^{\uparrow}$+ $p$ data at $\sqrt{s} =$ 510 GeV collected in 2017 at STAR. This dataset allows for a substantial enhancement in statistical precision.
\end{Abstract}
\vfill
\begin{Presented}
DIS2023: XXX International Workshop on Deep-Inelastic Scattering and
Related Subjects, \\
Michigan State University, USA, 27-31 March 2023 \\
     \includegraphics[width=9cm]{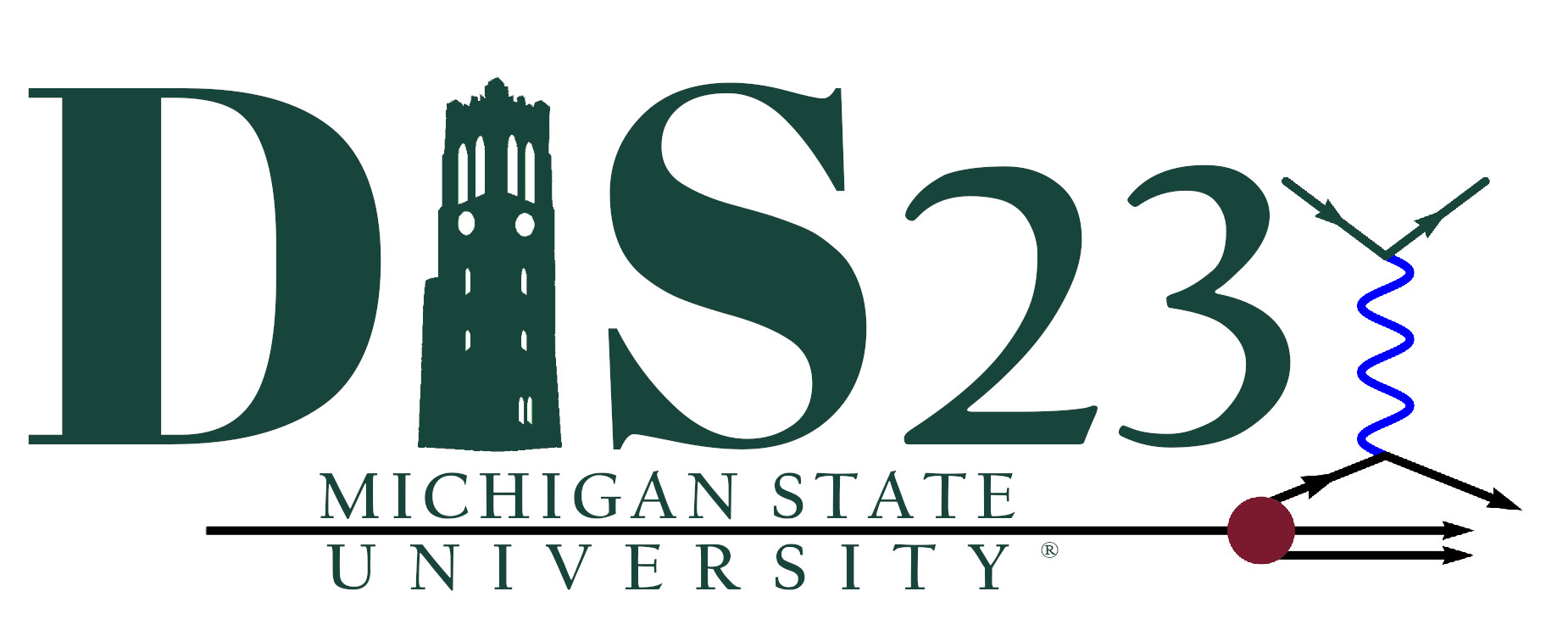}
\end{Presented}
\vfill
\end{titlepage}


\section{Introduction}
Transverse single-spin asymmetry, denoted by $A_{N}$, is also known as the left-right asymmetry of the particles produced with respect to the plane defined by the momentum and spin directions of the polarized beam. In recent decades, this asymmetry has been observed to be large for charged- and neutral-hadron production in polarized hadron-hadron collisions \cite{LargeTSSA1,LargeTSSA2,LargeTSSA3,LargeTSSA4,Zhanwen}. These observations stand in contrast to nearly zero $A_{N}$ predicted by perturbative Quantum Chromodynamics in the hard scattering processes \cite{pQCD}. Two major frameworks provide potential explanations for such sizeable asymmetries. The first one introduces the transverse-momentum-dependent contributions from the initial-state quark and gluon Sivers functions and/or the final-state Collins fragmentation functions \cite{Sivers, Collins}. The Sivers effect shows that this asymmetry comes from the correlation between the proton spin and the parton's transverse momentum at the initial state \cite{Sivers};  while the Collins effect arises from the correlation between the spin of the fragmenting quark and the transverse momentum of the resulting hadron at the final state \cite{Collins}. The second framework is based on the twist-3 contributions in the collinear factorization framework, which includes the contributions from the quark-gluon or gluon-gluon correlations and fragmentation functions \cite{Twist-3}. Additionally, experimental measurements indicate that the significant $A_{N}$ might arise from diffractive processes, according to the analyses of $A_{N}$ for forward $\pi^{0}$ and electromagnetic jets (EM-jets) in transversely polarized proton-proton (${p^{\uparrow}+p}$) collisions at STAR \cite{Zhanwen, Miganka}. 

In this proceeding, firstly, we present the preliminary results of $A_{N}$ for inclusive EM-jets in $p^{\uparrow}$+ $p$ collisions at $\sqrt{s} =$ 200 GeV based on the STAR 2015 dataset. These results explore the dependence of $A_{N}$ on photon multiplicity, transverse momentum ($p_{T}$), and energy of the EM-jets. Furthermore, we present the preliminary result for $A_{N}$ of diffractive EM-jets using ${p^{\uparrow}+p}$ collisions at $\sqrt{s} =$ 200 GeV from the same dataset. Finally, we show the statistical projection plots for $A_{N}$ of inclusive and diffractive EM-jets using ${p^{\uparrow}+p}$ collisions at $\sqrt{s} =$ 510 GeV from STAR 2017 data.

\section{Analysis}
\subsection{Experiment setup}
The measurements are conducted with the STAR experiment at the Relativistic Heavy Ion Collider (RHIC) at Brookhaven National Laboratory. RHIC is the only polarized proton-proton collider in the world, which is able to provide transversely or longitudinally polarized proton-proton collisions at $\sqrt{s} =$ 200 GeV and 500/510 GeV. The presented measurements and statistical projections are performed using high luminosity datasets with transversely polarized $p^{\uparrow}$ + $p$ collisions at $\sqrt{s} =$ 200 GeV and 510 GeV, respectively. Their average beam polarizations are about 57\% and 55\%, and their integrated luminosities are about 52 $\mathrm{pb}^{-1}$ and 350 $\mathrm{pb}^{-1}$, respectively. 

The major detectors used for these analyses are the Forward Meson Spectrometer (FMS) and the Roman Pot (RP) detectors. The FMS serves as an electromagnetic calorimeter designed to detect photons, neutral pions, and $\eta$ mesons. Located on the west side of the main STAR apparatus and about 7 meters away from the nominal interaction point, the FMS offers full azimuthal coverage and a pseudo-rapidity range of 2.6 to 4.2 \cite{FMS}. The RP detectors are located on both sides, about 15.8 meters from the nominal interaction point along the beamline. Each side features two sets of RP detectors, separated by approximately 1.8 meters. Within each set, there is a package with 4 silicon strip detector planes (SSDs) located above and below the beamline \cite{RP1}.

\subsection{Electromagnetic jet reconstruction and corrections}
\label{EM-jets}

The EM-jet is the EM component of a full jet. To reconstruct the EM-jets, first, the FMS clusters were formed by grouping adjacent towers with non-zero energies. Then, a shower shape fitting was performed for every cluster to obtain the FMS points as the photon candidates, which were used in EM-jet reconstruction for the analyses. Further information regarding the FMS photon candidates can be found in \cite{Zhanwen}. The anti-$k_{T}$ algorithm was employed to reconstruct the EM-jets, with a resolution parameter of $R = 0.7$  \cite{FASTJET}. The minimum $p_{T}$ requirement for the EM-jets was determined by either the trigger threshold or a fixed threshold depending on the dataset being analyzed.

The reconstructed EM-jet energy and $p_{T}$ were first corrected by subtracting the contribution from the underlying event, which was estimated using the ``off-axis'' cone method \cite{UE}. In addition, the EM-jet kinematics were further corrected back to the ``particle level'' based on the simulation, in order to account for the detector response. This simulation framework was set up with PYTHIA 6 with Perugia 2012 Tune for the particle level event generation \cite{Pythia,Tune2012}. The generated events were then passed through the GEANT-based STAR detector simulation.

\subsection{Channels and event selection for inclusive and diffractive processes}
The channels through which inclusive EM-jets are studied are $p^{\uparrow} + p \rightarrow \text{EM-jet} + X$. 

The presence of the rapidity gap between the RP and the FMS fulfilled the requirement for the diffractive processes. Consequently, diffractive events were identified by tagging the proton detected by the RP and identifying the EM-jets from the FMS. Two possible channels for the diffractive processes were considered: $p^{\uparrow} + p \rightarrow$ $p + \text{EM-jet} + X$ and $p^{\uparrow} + p \rightarrow$ $p + p + \text{EM-jet} + X$. Both channels required exactly one proton detected in the RP on the west side. The former channel required no proton detected on the east side, while the latter required exactly one proton detected on the east side.

The EM-jet reconstruction and correction procedures for inclusive processes and diffractive processes followed the methodology mentioned in the previous section \ref{EM-jets}. Additional event selection criteria were applied to identify the diffractive events. Firstly, the number of tracks detected in the RP (RP track) had to match the expected number of protons for the either possible channel of the diffractive processes. Moreover, these RP tracks are required to reconstruct properly based on the geometric acceptance of the RP. Then, the sum of the energy from the west side RP track and EM-jets, referred to as sum energy, are not allowed to exceed the threshold. Finally, the cut based on the ADC value of Beam-Beam Counter (BBC) \cite{BBC} was employed. Only the events with BBC ADC values not exceeding the specified threshold were retained. These final two cuts are able to reduce the fraction of background events significantly. More comprehensive information on these event selection criteria can be found in \cite{DIS 2022 proceeding}. 

\section{Results}
\subsection{Analysis method}
The cross-ratio method was used to extract the $A_{N}$ for both inclusive and diffractive processes, and the corresponding formulas are presented in Eq.~\ref{e1}  and \ref{e2}. In both equations, $A_{raw}$ represents the raw asymmetry obtained from the yields $N^{\uparrow (\downarrow)}(\phi)$ , $N^{\uparrow (\downarrow)}(\phi + \pi)$  observed at azimuthal angle $\phi$, $(\phi + \pi)$ relative to the polarized beam direction for spin up (down) state. The term $P$ corresponds to the average polarization of the proton beam. The cosine fit was applied to extract the $A_{N}$ from the raw asymmetry in Eq.~\ref{e2}.   

\begin{equation}
    \label{e1}
    A_{raw}(\phi)  = \frac{\sqrt{N^{\uparrow} (\phi)N^{\downarrow} (\phi + \pi)} - \sqrt{N^{\downarrow} (\phi)N^{\uparrow} (\phi + \pi)}}{\sqrt{N^{\uparrow} (\phi)N^{\downarrow} (\phi + \pi)} + \sqrt{N^{\downarrow} (\phi)N^{\uparrow} (\phi + \pi)}}
    \label{e1}
\end{equation}

\begin{equation}
    A_{raw}(\phi) = P A_{N} cos(\phi)
    \label{e2}
\end{equation}

This method takes advantage of detector azimuthal symmetry and cancels effects from the non-uniform detector efficiency and luminosity.

\subsection{Inclusive EM-jet $A_{N}$ for $p^{\uparrow}$+ $p$ data at $\sqrt{s} =$ 200 GeV}

\begin{wrapfigure}{R}{0.6\textwidth}    
    \centering
    \includegraphics[width=0.6\textwidth]{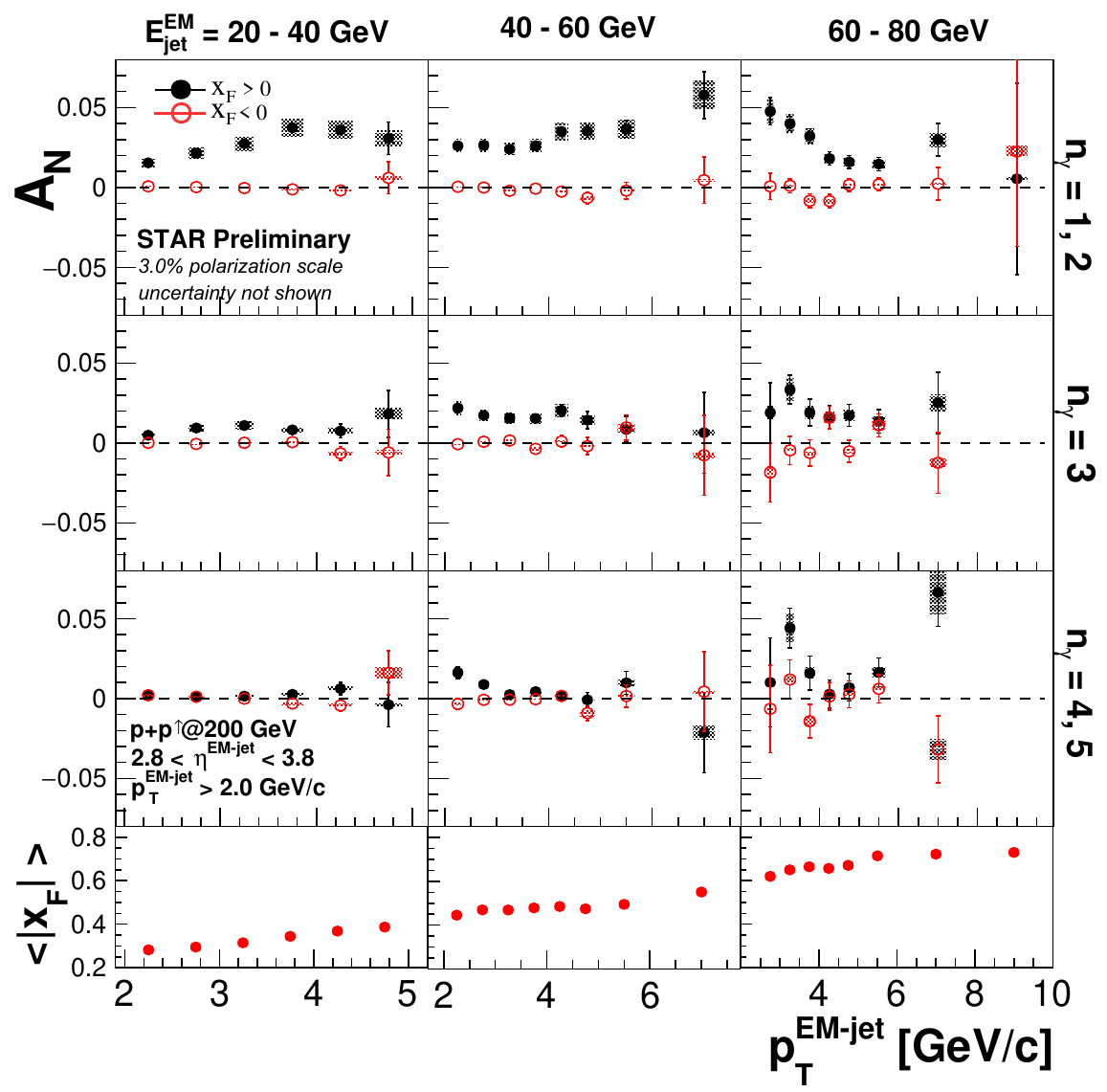}
    \caption{$A_{N}$ of inclusive EM-jet at FMS sorted by photon multiplicity, $p_{T}$, and energy bins. The lowermost panels display the average $x_{F}$ values corresponding to each $p_{T}$ bin. The black solid points represent the $A_{N}$ values for  $x_{F} > 0$ and the red hollow points depict the $A_{N}$ values for $x_{F} < 0$.}
    \vspace{-50pt}
    \label{Run 15 Inclusive EM-jet AN 1}
\end{wrapfigure}

Figure \ref{Run 15 Inclusive EM-jet AN 1} presents the preliminary results of the inclusive EM-jet $A_{N}$ as a function of photon multiplicity, EM-jet $p_{T}$, and EM-jet energy.  The $A_{N}$ decreases as the photon multiplicity of the EM-jets increases. Notably, the EM-jets consisting of 1 or 2 photons exhibit the most pronounced asymmetry.  The $A_{N}$ for $x_{F} < 0$ ($x_{F}$ is the longitudinal momentum fraction $x_{F} = 2p_{L}/\sqrt{s}$) is found to be consistent with zero regardless of the photon multiplicity. 


In addition, the photon multiplicity dependent inclusive EM-jet $A_{N}$ as a function of $x_{F}$ are presented in Fig.~\ref{Run 15 Inclusive EM-jet AN 2 and diffractive} (left). The inclusive EM-jet $A_{N}$ exhibits an increasing trend as $x_{F}$ increases, regardless of the photon multiplicity. Also, the $A_{N}$ of the EM-jet consisting of 1 or 2 photons is the strongest. This finding aligns with the previous measurement at STAR, where the $A_{N}$ of the isolated $\pi^{0}$ was observed to be higher than that of the non-isolated $\pi^{0}$ \cite{Zhanwen}.

\subsection{Diffractive EM-jet $A_{N}$ for $p^{\uparrow}$+ $p$ data at $\sqrt{s} =$ 200 GeV}

\begin{figure}[tp]
    \centering
    \begin{subfigure}{0.55\textwidth}
        \includegraphics[width=\textwidth]{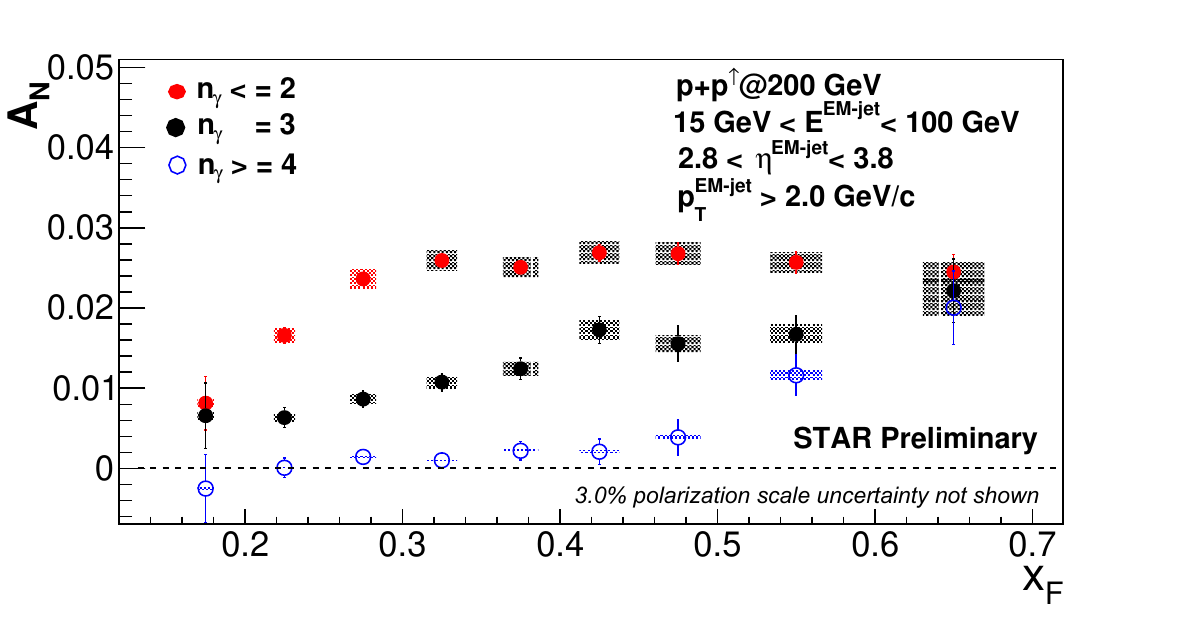}
    \end{subfigure}
    \hfill
    \begin{subfigure}{0.4\textwidth}

        \includegraphics[width=\textwidth]{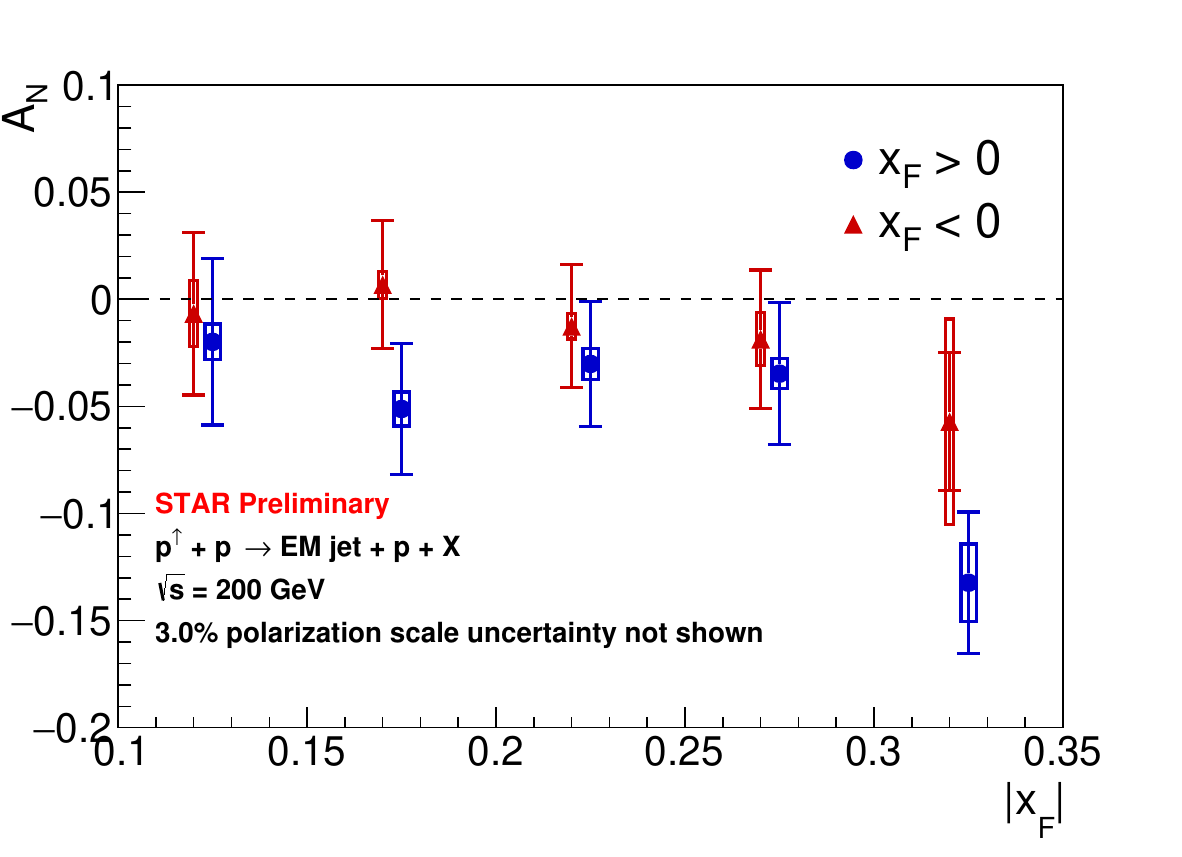}
    \end{subfigure}    
    \caption{(left) Inclusive EM-jet $A_{N}$ as a function of $x_{F}$ at $\sqrt{s} =$ 200 GeV for three cases: $n_{\gamma} \leq 2$, $n_{\gamma} = 3$, $n_{\gamma} \geq 4$. (right) Diffractive EM-jet $A_{N}$ as a function of $x_{F}$ at $\sqrt{s} =$ 200 GeV. The blue points represent $x_{F} > 0$, while the red points represent $x_{F} < 0$ with a constant shift of -0.005 along x-axis for clarity. The rightmost points correspond to $0.3 < |x_{F}| < 0.45$.}
    \label{Run 15 Inclusive EM-jet AN 2 and diffractive}
\end{figure}

Figure \ref{Run 15 Inclusive EM-jet AN 2 and diffractive} (right) presents the preliminary result for diffractive EM-jet $A_{N}$ as a function of $x_{F}$. We observe a non-zero diffractive EM-jet $A_{N}$ with a significance of 3.3$\sigma$ below 0 at forward rapidity. Moreover, a significant absolute $A_{N}$ is observed at the high $x_{F}$ region. However, the sign of the diffractive EM-jet $A_{N}$ is negative, which stands in contrast to the inclusive EM-jet $A_{N}$ in Fig.~\ref{Run 15 Inclusive EM-jet AN 1} and \ref{Run 15 Inclusive EM-jet AN 2 and diffractive} (left). The $A_{N}$ for $x_{F} < 0$ is found to be consistent with zero.  More theoretical inputs are needed to understand the behavior observed in the diffractive results.

\subsection{Statistical projection for $p^{\uparrow}$+ $p$ data at $\sqrt{s} =$ 510 GeV}
\begin{figure}[htbp]

    \centering
        \includegraphics[scale=0.55]{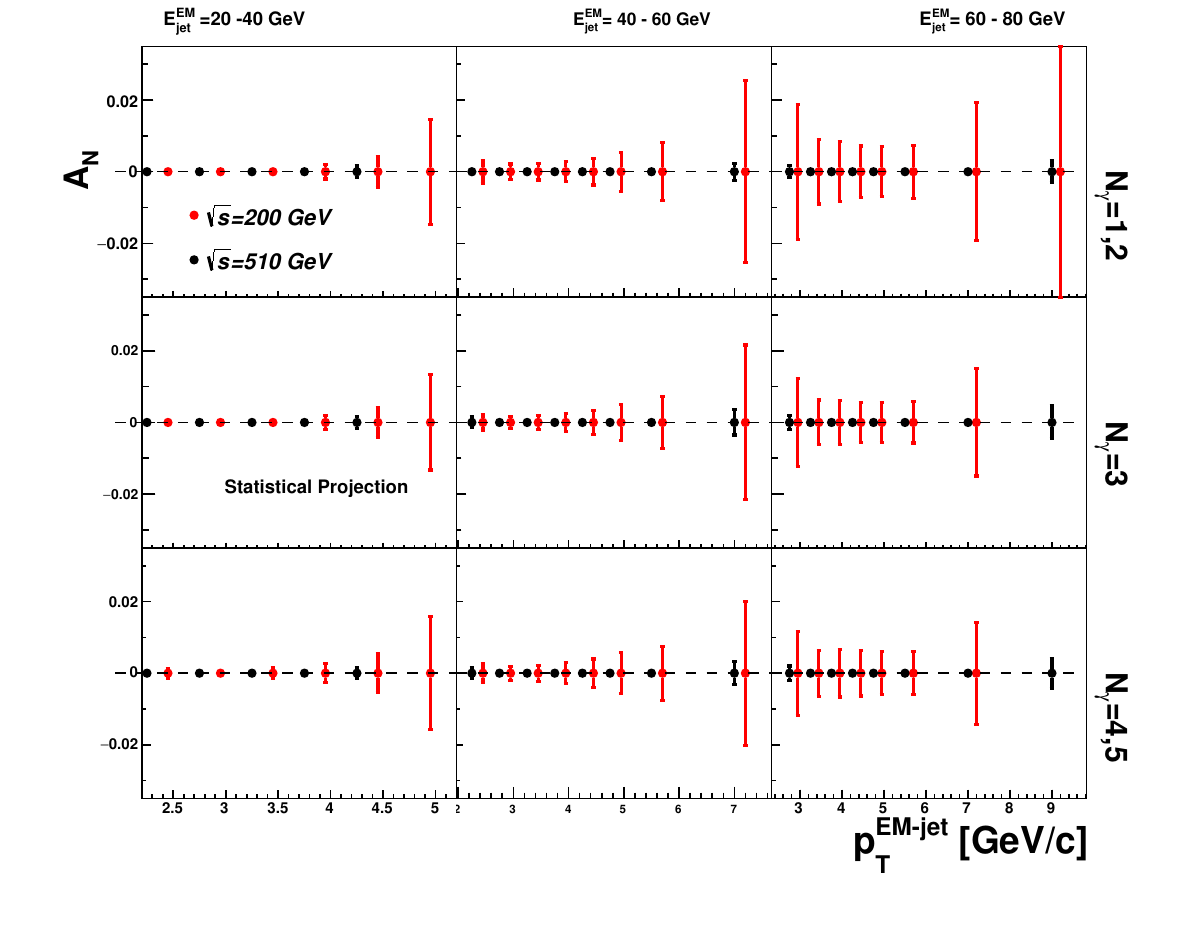}
        \caption{Statistical projections for inclusive processes  for $p^{\uparrow} + p$ collisions at $\sqrt{s}=$ 510 GeV (black) compared to results at $\sqrt{s}=$ 200 GeV (red) at STAR.}
    \label{Statistical projection}
\end{figure}

\begin{figure}[tbp]  
    \vspace{-15pt}
    \centering
    \includegraphics[width=0.5\linewidth]{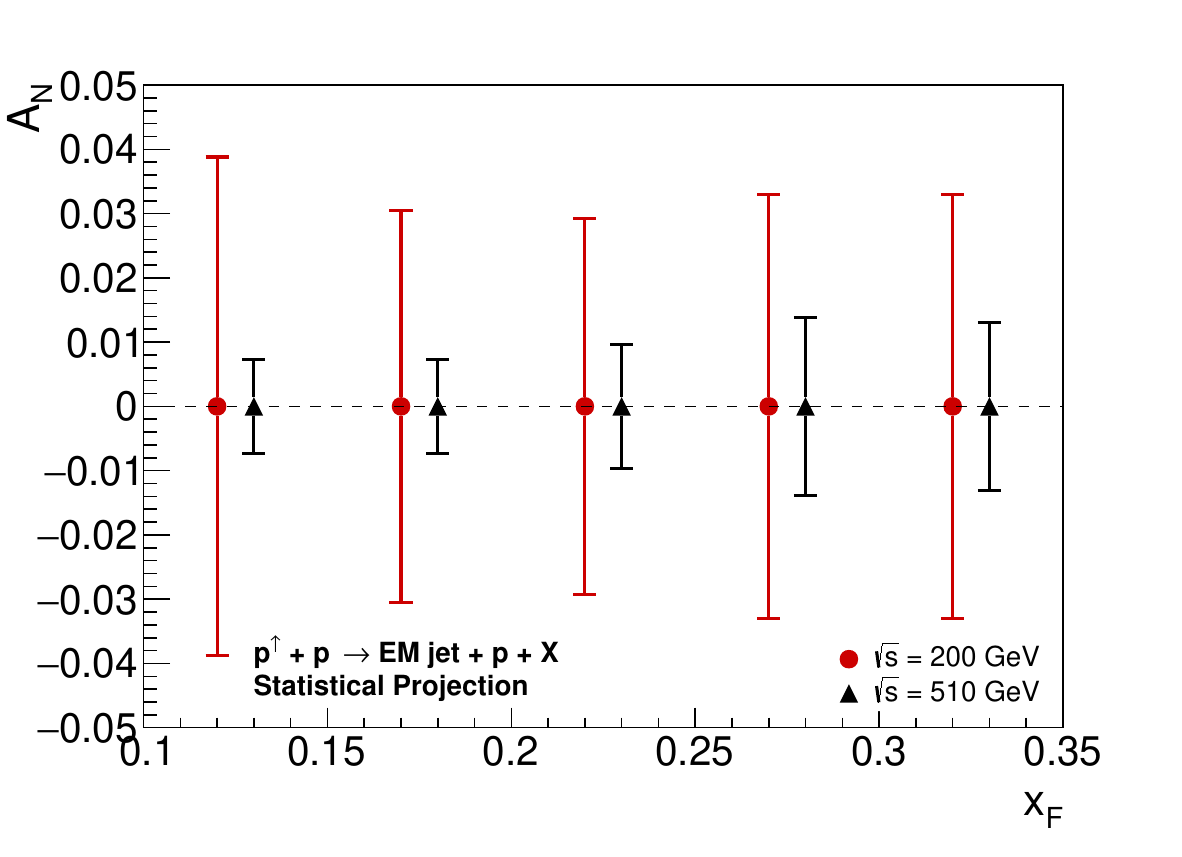}
    \caption{Statistical projections for diffractive processes for $p^{\uparrow} + p$ collisions at $\sqrt{s}=$ 510 GeV (black) compared to results at $\sqrt{s}=$ 200 GeV (red) at STAR.}
    \label{Sta diff}
    \vspace{-10pt}
\end{figure}

The ongoing analyses of $A_{N}$ for both inclusive and diffractive processes are being conducted using data at $\sqrt{s} =$ 510 GeV.  This high luminosity dataset holds promising prospects for a more precise investigation of $A_{N}$ in both inclusive and diffractive measurements. To illustrate the anticipated improvements, Fig.~\ref{Statistical projection}  shows the statistical projection for the inclusive processes, while Fig.~\ref{Sta diff}  presents the statistical projection for the diffractive processes. These plots compare the data at $\sqrt{s} =$ 200 GeV and 510 GeV. With the utilization of the $\sqrt{s} =$ 510 GeV data, a significant improvement in the precision of $A_{N}$ measurements is expected, resulting in a reduction in statistical uncertainty of about a factor of 3 for high energy and high photon multiplicity EM-jets for inclusive EM-jet $A_{N}$ measurement, and more than a factor of 2 for diffractive EM-jet $A_{N}$ measurement.

\section{Conclusion}
We present the inclusive and diffractive EM-jet $A_{N}$ using the FMS at STAR in $p^{\uparrow} + p$ collisions at $\sqrt{s}=$ 200 GeV. The $A_{N}$ for inclusive EM-jets increased with  $x_{F}$. Notably, the $A_{N}$ with lower photon multiplicity for the inclusive processes was found to be larger. The $A_{N}$ for the diffractive processes is non-zero with a significance of 3.3 $\sigma$. However, the sign of diffractive $A_{N}$ is negative, which is opposite to that observed in the inclusive processes. Further theoretical inputs are needed to understand its underlying physics. Finally, with the higher luminosity data set for $p^{\uparrow} + p$ collisions at $\sqrt{s}=$ 510 GeV at STAR, a higher precision will be achieved for both the inclusive and diffractive EM-jet $A_{N}$.

\end{document}